\documentclass[twocolumn, superscriptaddress]{revtex4-1}

\usepackage[utf8]{inputenc}
\usepackage[english]{babel}
\usepackage[T1]{fontenc}

\usepackage{graphicx}
\usepackage{float}
\usepackage{xcolor}
\usepackage{braket}

\usepackage{amsmath,amsfonts,amssymb}

\usepackage{ntheorem}
  \theoremseparator{.}
  
  \newtheorem{definition}{Definition}

\usepackage{hyperref}

\DeclareMathOperator{\tr}{tr}

\newcommand{\hilb}{\mathcal{H}}

\newcommand{\vm}[1]{\left\langle{#1}\right\rangle}

\begin{document}
\title{Characterization of non-signaling correlations from mutual information}

\author{Ignacio Perito}
\affiliation{Departamento de Física, Facultad de Ciencias Exactas y Naturales, Universidad de Buenos Aires, 1428 Buenos Aires, Argentina}
\affiliation{CONICET-UBA, Instituto de Física de Buenos Aires (IFIBA), 1428 Buenos Aires, Argentina}
\author{Guido Bellomo}
\affiliation{Departamento de Física, Facultad de Ciencias Exactas y Naturales, Universidad de Buenos Aires, 1428 Buenos Aires, Argentina}
\affiliation{CONICET-UBA, Instituto de Investigación en Ciencias de la Computación (ICC), 1428 Buenos Aires, Argentina}
\author{Daniel Galicer}
\affiliation{Departamento de Matemática, Facultad de Ciencias Exactas y Naturales, Universidad de Buenos Aires, 1428 Buenos Aires, Argentina, and IMAS-CONICET}
\author{Santiago Figueira}
\affiliation{CONICET-UBA, Instituto de Investigación en Ciencias de la Computación (ICC), 1428 Buenos Aires, Argentina}
\affiliation{Departamento de Computación, Facultad de Ciencias Exactas y Naturales, Universidad de Buenos Aires, 1428 Buenos Aires, Argentina}
\author{Augusto J. Roncaglia}
\affiliation{Departamento de Física, Facultad de Ciencias Exactas y Naturales, Universidad de Buenos Aires, 1428 Buenos Aires, Argentina}
\affiliation{CONICET-UBA, Instituto de Física de Buenos Aires (IFIBA), 1428 Buenos Aires, Argentina}
\author{Ariel Bendersky}
\affiliation{CONICET-UBA, Instituto de Investigación en Ciencias de la Computación (ICC), 1428 Buenos Aires, Argentina}
\affiliation{Departamento de Computación, Facultad de Ciencias Exactas y Naturales, Universidad de Buenos Aires, 1428 Buenos Aires, Argentina}

\date{\today}

\begin{abstract}
We present a characterization of the set of non-signaling correlations in terms of a two dimensional representation that involves the maximal value of a Bell functional and the mutual information between the parties. In particular, we apply this representation to the bipartite Bell scenario with two measurements and two outcomes. In terms of these physically meaningful quantities and through numerical optimization methods and some analytical results, we investigate the frontier between the different subsets of the non-signaling correlations, focussing on the quantum and post-quantum ones. Finally, we show that the Tsirelson bound appears as a singular point in this context without resorting to quantum mechanics.
\end{abstract}

\maketitle
 
\section{Introduction}\label{sec:intro}

When a set of spatially separated systems is considered, the complete characterization  from physical principles of  the set of quantum correlations that can exist between is still an open problem. The simplest attempt to do so corresponds to the principle of no-signalization between the systems~\cite{cirel1980quantum,popescu1994quantum}, which is well known to give place to a set of correlations for which those achievable within the quantum mechanics formalism are just a subset. There are more sophisticated physical principles trying to define the frontier between quantum correlations and  correlations that cannot be obtained from quantum mechanics, these proposals include the principle of information causality~\cite{pawlowski2009information}; non-trivial communication complexity~\cite{cleve1997substituting}; no advantage for nonlocal computation~\cite{linden2007quantum}; macroscopic locality~\cite{navascues2010glance}; local orthogonality~\cite{fritz2013local} to name a few. Some of them are able to provide a good approximation to the set of quantum correlations or can even describe it exactly in some particular scenarios, but a general physical principle defining the quantum set in the general case is yet unknown and one of the main open problems in this research field.

In this work, we  present an alternative approach to the problem of characterizing quantum correlations using a two-dimensional representation of the non-signaling set. 
We focus on the bipartite scenario where two spatially separated parties have access to two dichotomic 
measurement choices each. This representation makes use of two real functionals acting on the set of correlations: the CHSH functional, which is a well known quantity, directly related to the geometry of the non-signaling set; and the mutual information between the parties, which is a faithful measure of the strength of the correlations that exist between them. We characterize the boundaries of the different sets of correlations and show the natural appearance of the Tsirelson bound as a singular point in this representation. Interestingly, this singular point emerges without assuming the quantum mechanics formalism.

The paper is organized as follows: in Sec.~\ref{sec:scenario} we define the scenario we will be using through out this work, discuss some previous results and provide some relevant definitions. Sec.~\ref{sec:results} contains our main results and is divided in three parts: in \ref{sec:sym} we discuss the particular case of symmetrical behaviors which are relevant for our approach; in \ref{sec:quantum} we give some insights about how quantum correlations are distributed on our two-dimensional representation and in \ref{sec:tsirelson} we show an interesting fact which suggests that the Tsirelson bound could be obtained in a device-independent manner. Finally, in Sec.~\ref{sec:summary} we summarise our results. 

\section{Scenario and Definitions}\label{sec:scenario}
\subsection{Non-signaling set}

We will  focus on the standard device-independent Bell scenario $(2,2,2)$: two parties, Alice and Bob, 
each of which has access to a device with two inputs (measurement choices) and  outputs. 
We note by $x\in\{0,1\}$ and $y\in\{0,1\}$ the measurement choices of Alice and Bob, respectively; and $a$ and $b$ their 
possible outcomes, which take values $\{-1,1\}$. This kind of scenario is completely described once the $16$ conditional probabilities $p(ab | xy)$ are given, which are usually referred as \textit{behaviors}~\cite{tsirelson1993some}. Each behavior can be represented by a vector with real components 
${\bf p} =\{p(ab | xy)\}$, that satisfies  the normalization condition
$\sum_{ab} p(ab|xy) = 1$  and 
the positivity constraints $p(ab|xy) \geq 0$. Thus, leaving an amount of $12$ independent parameters. 

Additional constraints can be imposed based on physical considerations. 
The first one defines  $\mathcal{NS}$, the \textit{non-signaling} set, with behaviors whose marginal probabilities are locally well defined \cite{cirel1980quantum,popescu1994quantum}:
\begin{equation} \label{eq:condns}
\begin{split}
  p( a| x ) &= \sum_{b}   p( ab | xy ) \quad \forall x, \\
p( b| y ) &= \sum_{a}   p( ab | xy ) \quad \forall y \, .
\end{split}
\end{equation}
These conditions guarantee that instantaneous communication between the parties is forbidden, which is required to preserve the causality principle. The $\mathcal{NS}$ set in the $(2,2,2)$ scenario is then an {8-dimensional} polytope embedded in $\mathbb{R}^{16}$. 
In this case, instead of specifying a behaviour in terms of the 16 components, a particularly simple parametrization of $\mathcal{NS}$ is given in terms 
of 8 \textit{correlators} 
$\{\vm{A_x},\vm{B_y},\vm{A_x B_y} \}$:
\begin{equation} \label{eq:meanvalues}
\begin{split}
\vm{A_x}  \equiv \sum_a a\, p(a|x), &\quad \quad
\vm{B_y} \equiv \sum_b b\, p(b|y), \\
 \vm{A_xB_y} &\equiv \sum_{ab} ab \ p(ab|xy),
\end{split}
\end{equation}
from which any behavior can be obtained as \cite{brunner2014bell}:
\begin{equation}  \label{eq:vmtop}
\begin{split}
&p(ab|xy) = \frac{1}{4}(1+a\vm{A_x} + b\vm{B_y} + ab\vm{A_xB_y}),
\end{split}
\end{equation}
and using \eqref{eq:condns} we can also obtain the marginals in terms of the correlators. It is easy to check that all the mean values defined above are real numbers lying in $[-1,1]$. Therefore, the $\mathcal{NS}$ set can also be described as the intersection of the hypercube $[-1,1]^8 \subset \mathbb{R} ^8$ and the $16$ semi-spaces defined by the positivity conditions. 

Let us introduce two interesting subsets of $\mathcal{NS}$ that will be useful for our analysis. The first one is the well-known \textit{correlation space} $\mathcal{C}$ \cite{cirel1980quantum,avis2009bell}, defined by the set of behaviors with $\vm{A_x} = \vm{B_y} = 0$ for all $x$ and $y$. This set describes devices that work as perfectly unbiased coins with no dependence on the inputs. Notice also that in this case the positivity conditions are automatically satisfied, so the correlation space is a {4-dimensional} subset of $\mathcal{NS}$ in one-to-one correspondence with all the points in the hypercube $[-1,1]^4$. The other subset of $\mathcal{NS}$   is the set $\mathcal{SYM}$ of \textit{symmetric behaviors}, corresponding to those elements of $\mathcal{NS}$ that are invariant under  exchange of the devices. Formally, it is obtained by imposing three additional constraints to the correlators that parametrize the non-signaling polytope:
$\vm{A_x} = \vm{B_y}$ for $x=y$ and 
$\vm{A_0B_{1}}= \vm{A_1B_{0}}$
which makes $\mathcal{SYM}$ a 5-dimensional polytope.

Another subset of particular physical interest is the set $\mathcal{L}$ of \textit{local behaviors}, which contains all the behaviors admitting a local model:
\begin{equation}
p(ab|xy) = \int_{\Lambda} d\lambda \  p(a|x,\lambda) \ p(b|y,\lambda)  \, ,
\end{equation}
where $\lambda$ are arbitrary variables taking values in some set   $\Lambda$ shared by both devices, usually refer as hidden variables. This set is a polytope that is strictly smaller than the non-signaling set, that is $\mathcal{L}\subset \mathcal{NS}$. Its vertices are known as the local deterministic behaviors (\textit{LD behavior} in the following), and correspond to those situations ($16$ in our scenario) where both devices work as deterministic functions of their inputs. The behaviors that do not belong to this set are called \emph{nonlocal}.

The last physically relevant set is $\mathcal{Q}$, the \emph{quantum set}. This is the set of behaviors allowed by quantum mechanics. A behavior belongs to $\mathcal{Q}$ if and only if there exist two Hilbert spaces $\hilb_A$ y $\hilb_B$, and a density matrix $\rho \in \hilb_A \otimes \hilb_B$ of arbitrary dimension
such that:
\begin{equation}\label{eq:quantbeh}
p(ab|xy) = \tr \left[ \rho \left( M_{a|x} \otimes M_{b|y} \right) \right] \, ,
\end{equation}
where $\{ M_{a|x} \}$ is a set of measurement operators (all $M_{a|x}$ are semi-definite positive and $\sum_{a}M_{a|x} = \mathbb{I}_A$) and analogously $\{M_{b|y} \}$  for Bob's measurements. Note that, by means of purification, we can assume without loss of generality that the density matrix corresponds to a pure state and the measurements are orthogonal projectors, so any element of $\mathcal{Q}$ can also be written as:
\begin{equation}\label{eq:quantbehpurif}
p(ab|xy) = \bra{\psi} M_{a|x} \otimes M_{b|y} \ket{\psi} \, ,
\end{equation}
where now $\ket{\psi}$ is a pure quantum state in $\hilb_A \otimes \hilb_B$ and all the involved measurements are projective: $M_{a|x}M_{a'|x}=\delta_{aa'}M_{a|x}$,
and similarly for Bob's measurements. 
The local and quantum sets have the same dimension as the non-signaling set \cite{pironio2005lifting} and  satisfy the following strict inclusions: $\mathcal{L} \subset \mathcal{Q} \subset \mathcal{NS}$ \cite{brunner2014bell}.  

The distinction between the quantum behaviors and the post-quantum ones  (those behaviors that are non-signaling but lie outside the quantum set) is clear from an operational point of view (a behavior is quantum if and only if it can be written as \eqref{eq:quantbehpurif}). But this distinction is not yet fully understood in terms of a physical point of view, and in this work we will present an informational approach to this problem. In order to do so, we will make use of two functionals that map behaviors into real numbers. Before doing so, a little digression about the geometry of the three sets \cite{brunner2014bell} will come in handy. As we said, using the parametrization in terms of correlators, both normalization and non-signaling constraints are automatically fulfilled. Therefore, the non-signaling polytope is defined just by specifying its $16$ facets, given by the positivity constraints. Local behaviors can also attain the bound imposed by the positivity conditions, so there are $16$ facets of $\mathcal{L}$ that are contained in the facets of $\mathcal{NS}$. Given that $\mathcal{L} \subset \mathcal{Q} \subset \mathcal{NS}$, the last assertion must also hold for $\mathcal{Q}$ (which is convex but not a polytope so not all of its boundary will be given by hyperplanes). The inner facets of $\mathcal{L}$, that is, the facets that do not correspond to positivity constraints, establish the frontier between local and nonlocal behaviors. Being hyperplanes, they can be described by level surfaces of linear functionals, called Bell inequalities, acting over the set of behaviors. For instance, the well-known CHSH 
functional
\cite{clauser1969proposed}:
\begin{equation} \label{eq:chshgen}
S \equiv \vm{A_0B_0} + \vm{A_0B_1} + \vm{A_1B_0} - \vm{A_1B_1} \,,
\end{equation}
is such  that local behaviors must satisfy $-2 \leq S\leq 2$, so the two hyperplanes $S = \pm 2$ are facets of $\mathcal{L}$. The quantum bound for $S$ can be easily obtained as $-2\sqrt{2} \leq S\leq 2\sqrt{2}$ and  is known as Tsirelson bound. 
In this case,  $\mathcal L$ is defined by the eight inequalities obtained by relabeling inputs on $-2 \leq S\leq 2$, along with the positivity constraints.
In the following, we refer to behaviors attaining Tsirelson bound as \textit{Bell behaviors}. 

The quantum set $\mathcal{Q}$  is convex but it is not a polytope, it is bounded by some  non-flat regions and any set of inequalities describing it must necessarily be non-linear. A closed set of expressions defining whether a behavior is quantum or not is not known in the general case, but there exist useful approximations to bound the quantum set. For example, in the present scenario, a necessary condition for a behavior to be quantum is \cite{landau1988empirical,tsirelson1993some,masanes2005extremal}:
\begin{equation} \label{eq:Qhatcond}
\left| \sum_{x'y'} \arcsin\vm{A_{x'}B_{y'}} - 2 \arcsin \vm{A_xB_y} \right| \leq \pi \quad \forall x,y \, 
\end{equation}
which, notably, when considering behaviors in the correlation space becomes also a sufficient condition. That is, the set $\mathcal{Q} \cap \mathcal{C}$ is completely described by the set of four inequalities \eqref{eq:Qhatcond}. In the general case, the set satisfying \eqref{eq:Qhatcond}, which we will call $\tilde{\mathcal{Q}}$ contains the quantum set $\mathcal{Q}$. A better approximation to the quantum set \cite{navascues2008convergent} is given by those behaviors where at least one $x$ (or $y$) such that $ \vm{A_x} = \pm1$ (or $\vm{B_y} = \pm1$); or:
\begin{equation}\label{eq:NPA1cond}
\left| \sum_{x'y'} \arcsin \left( \mathcal{F}_{x'y'} \right) - 2 \arcsin \left( \mathcal{F}_{xy} \right)  \right| \leq \pi
\end{equation}   
for all $x,y$, with:
\begin{equation}
\mathcal{F}_{xy} \equiv \frac{\vm{A_xB_y}-\vm{A_x}\vm{B_y}}{\sqrt{\left(1-\vm{A_x}^2\right)\left(1-\vm{B_y}^2\right)}} \, .
\end{equation}
A behavior satisfying these conditions is said to belong to the first level of the NPA hierarchy \cite{navascues2007bounding}, and we will refer to it as $\mathcal{Q}_1$. In general, then, we have $\mathcal{Q} \subset \mathcal{Q}_1 \subset \tilde{\mathcal{Q}}$.  

\begin{figure}
    \centering
    \includegraphics[width=0.8\columnwidth]{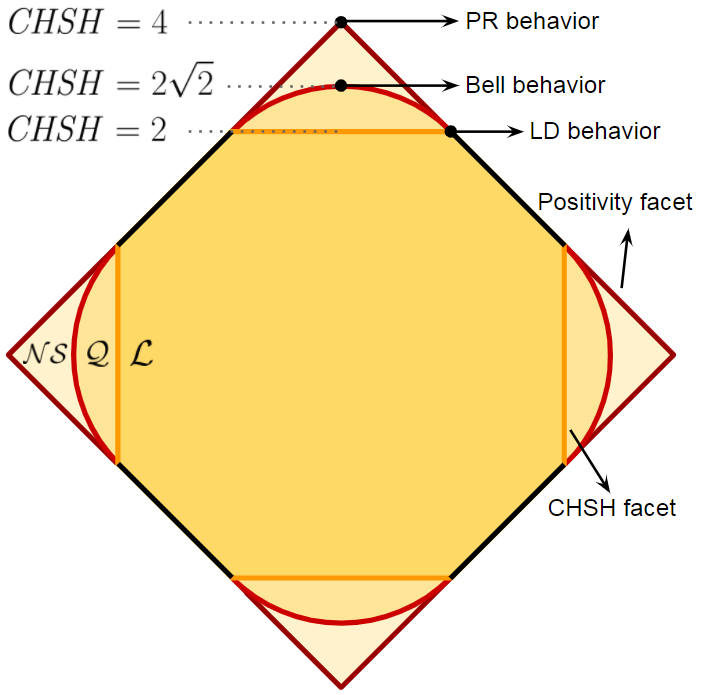}
    \caption{Pictorial representation of the non-signaling set. Dark red lines represent positivity constraints and orange lines represent CHSH facets. The black segments are those portions of the positivity constraints that are also boundaries of $\mathcal{L}$ and $\mathcal{Q}$. The non-flat boundaries of $\mathcal{Q}$ are shown in red.}
    \label{fig:sets}
\end{figure}

The algebraic bounds for $S$ are clearly $-4 \leq S \leq 4$ and, notably, can be attained by elements on $\mathcal{NS}$. This kind of behaviors are associated to devices called PR boxes \cite{popescu1994quantum}, and they correspond to the situations in which for instance:
$\vm{A_0B_0} = \vm{A_0B_1} = \vm{A_1B_0} = 1 = - \vm{A_1B_1}$.
We will refer to this kind of nonlocal behaviors as \textit{PR behaviors}. 
As a summary, Fig.~\ref{fig:sets} schematizes the geometric structure of the non-signaling
polytope we have discussed in the last paragraphs. 

\subsection{Functionals}

Let us define the two main functionals over which our analysis will be based. The first one, was already mentioned in a sense and is related to the CHSH functional: it is the maximum of such quantity over all possible relabelings of inputs and outputs. 

\begin{definition}
Let $\mathbf{p} \in \mathcal{NS}$ de an arbitrary non-signaling behavior. The functional $\mathcal{S}: \mathcal{NS} \rightarrow \mathbb{R}$ is defined as:
\begin{equation}
\mathcal{S} [\mathbf{p}] \equiv \max_{xy} \left| \left( \sum_{x'y'} \vm{A_{x'}B_{y'}} \right) - 2 \vm{A_x B_y} \right| \, .
\end{equation} 
\end{definition}

Maximizing over all CHSH functionals is important because it gives us a quantity that does not depend upon particular relabelings of inputs and/or outputs, which have no physical relevance. This is so because for any nonlocal behavior, there is always a relabeling for which the value of a particular CHSH functional does not violate the local bound. In other words, a behavior $\mathbf{p} \in \mathcal{NS}$ is nonlocal if and only if $\mathcal{S}[\mathbf{p}] > 2$. Behaviors satisfying $\mathcal{S} [ \mathbf{p} ] = 2$ lay at one of the non-trivial facets of the local set, and any deviation from this value indicates how far it is from its closest non-trivial facet of $\mathcal{L}$.

The second functional we will consider in this work is related to the correlations between the distribution of outputs of the devices. The mutual information is a faithful measure of the strength of the correlations between variables.
The mutual information between two random variables $W$ and $R$, conditioned to a third random variable $T$ can be computed as \cite{cover1999elements}:
\begin{equation} \label{eq:infomutuagen}
I ( W;R | T) = H(W|T) + H(R|T) - H(WR|T) \, ,
\end{equation}
where $H(\star | \star)$ is the conditional entropy. In our scenario, if we refer $X$ and $Y$ to the random variables associated with the inputs of both devices, and $A$ and $B$ to the random variables associated with the outputs, the mutual information between Alice's and Bob's variables can be written as:
\begin{equation}
I(A;B | XY) = H(A|X) + H(B|Y) - H(AB | XY).
\end{equation}
Where we have used the non-signaling conditions to simplify the first two terms as $H(A|XY) = H(A|X)$ and $H(B|XY) = H(B|Y)$. The conditional entropies appearing in \eqref{eq:infomutuagen} can be easily computed in terms of behaviors:
\begin{equation}
\begin{split}
H(A|X) &= \sum_{x} p(x) H(A | X=x) \\
&= - \sum_{ax} p(x) p(a|x) \log p(a|x) \, ,
\end{split}
\end{equation}
and similarly for $H(B|Y)$, and
\begin{equation}
\begin{split}
H(AB|XY)&= \sum_{xy} p(xy) H (AB|X=x,Y=y) \\
&= - \sum_{abxy} p(xy) p(ab|xy) \log p(ab|xy) \, .
\end{split}
\end{equation}
Naturally, as it is evident from the last two expressions, the mutual information depends on the probability distributions for the inputs. As we are interested in an intrinsic measure of the correlations contained in a given behavior, we will consider that both inputs are independent and uniformly distributed random variables.

\begin{definition}
Let $\mathbf{p} \in \mathcal{NS}$ be an arbitrary non-signaling behavior. We define the functional $\mathcal{I} : \mathcal{NS} \rightarrow \mathbb{R}$ as:
\begin{equation}
\begin{split}
\mathcal{I} [\mathbf{p}] \equiv& -\frac{1}{2} \sum_{ax} p(a|x) \log p(a|x)  \\ 
& -\frac{1}{2} \sum_{by} p(b|y) \log p(b|y) \\
& +\frac{1}{4} \sum_{abxy} p(ab|xy) \log p(ab|xy) \, ,
\end{split}
\end{equation}
\end{definition}

We will measure entropy in bits, which means that we will use base $2$ logarithms through this work. This implies that the functional $\mathcal{I}$ is a real number between $0$ (no correlation) and $1$ (maximum correlation between the parties). 

\begin{figure}
    \centering
    \includegraphics[width=1\columnwidth]{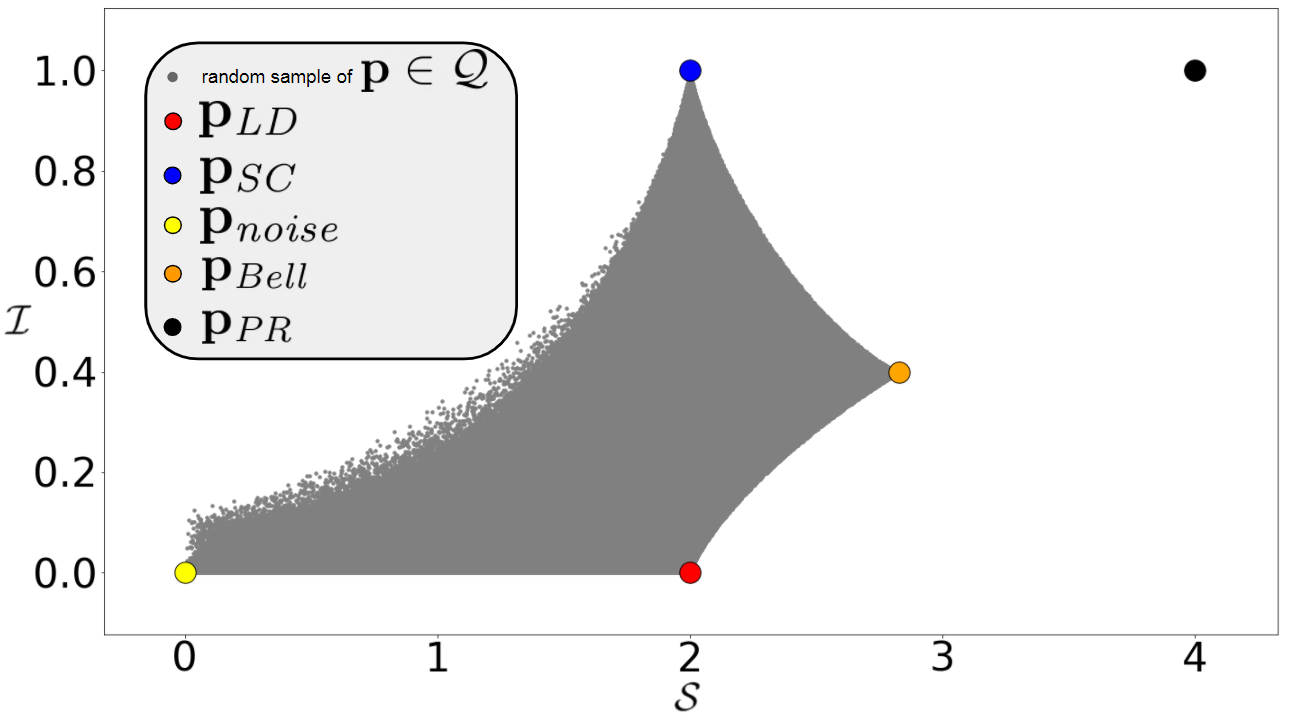}
    \caption{$\mathcal{I}$ vs. $\mathcal{S}$ plot for a random sample of $5\times 10^6$ quantum points for two-level systems on both sides. Some relevant behaviors are also plotted.}
    \label{fig:plot}
\end{figure}

Before presenting the results, it is instructive to look at a typical $\mathcal{I}$ vs. $\mathcal{S}$ plot. Fig.~\ref{fig:plot} shows the values of both functionals for a random sample of $5\times 10^6$ quantum behaviors of the form \eqref{eq:quantbehpurif} for 2-dimensional Hilbert spaces. The plot also highlights some relevant points: the PR behavior, the Bell behavior and a local deterministic behavior. In addition to those behaviors, we plot the one that maximizes $\mathcal{I}$ in the local set: 
$\vm{A_xB_y} = 1$ and $\vm{A_x} = \vm{B_y}=0$ for all $x,y$; that is, both devices output a perfectly 
correlated random and uniformly distributed bit (we will refer to this behavior, or any of it relabelings, as the 
\textit{shared coin behavior} or simply \textit{SC behavior}). 
The other highlighted behavior is what we will call the \textit{noise}, corresponding to $\vm{A_xB_y}= \vm{A_x} = \vm{B_y} = 0$ for all $x,y$ (that is, $p(ab|xy) = 1/4$ for all $a,b,x,y$). 

Throughout this paper, we will work extensively with the boundaries between the different sets in this type of plots. These boundaries correspond to the points that maximize or minimize $\mathcal{I}$ for each value of $\mathcal{S}$. We will use the notation $I^{\mathcal{A}}_{\mathit{max}}\left(\mathcal{S}\right)$ to refer to the maximum value of mutual information over the set $\mathcal{A}$ when the CHSH functional is fixed to $\mathcal{S}$, and  $I^{\mathcal{A}}_{\mathit{min}}\left(\mathcal{S}\right)$ for the minimum.

\section{Results}\label{sec:results}

Let us call $G: \mathcal{NS} \rightarrow \mathbb{R}^2$ the continuous function that maps behaviors to the $\mathcal{S} - \mathcal{I}$ plane: $G(\mathbf{p}) \equiv \left(\mathcal{S}(\mathbf{p}),\mathcal{I}(\mathbf{p}) \right)$. It is well known that continuous functions map path-connected sets into path-connected sets, so we have that the $\mathcal{S}-\mathcal{I}$ representation of $\mathcal{NS}$ is a path-connected set. Moreover, the latter is also simply connected (colloquially, it does not have empty regions inside of its outer boundaries). To see this, for each value of $s \in [0,4]$, we can take $\mathbf{p}_\mathit{min}$ and $\mathbf{p}_\mathit{max}$ to be two behaviors attaining the values of $I^{\mathcal{NS}}_{\mathit{min}}\left(s\right)$ and $I^{\mathcal{NS}}_{\mathit{max}}\left(s\right)$ respectively and, without loss of generality, we can assume that both behaviors attain the value of $\mathcal{S} = s$ for the same linear version of the CHSH functional. Thus, all convex combinations of those two behaviors have also the same value of the linear CHSH functional and, given that all other relabelled of it give smaller or equal values, the value of the non-linear functional $\mathcal{S}$ is also $s$ for all such convex combinations. Then, all behaviors corresponding to convex combinations of $\mathbf{p}_\mathit{min}$ and $\mathbf{p}_\mathit{max}$ lay in the vertical line given by $\mathcal{S} = s$ in the $\mathcal{S} - \mathcal{I}$  representation and, given that the mutual information is a continuous function, the vertical segment connecting $\mathbf{p}_\mathit{min}$ and $\mathbf{p}_\mathit{max}$ in the $\mathcal{S} - \mathcal{I}$ plane is filled by behaviors on $\mathcal{NS}$. Therefore, there are no empty regions inside of the outer boundaries of the $\mathcal{S} - \mathcal{I}$ representation of $\mathcal{NS}$ and, in order to characterize the non-signaling set, it is enough to find those outer boundaries. 

\begin{figure}
    \centering
    \includegraphics[width=1\columnwidth]{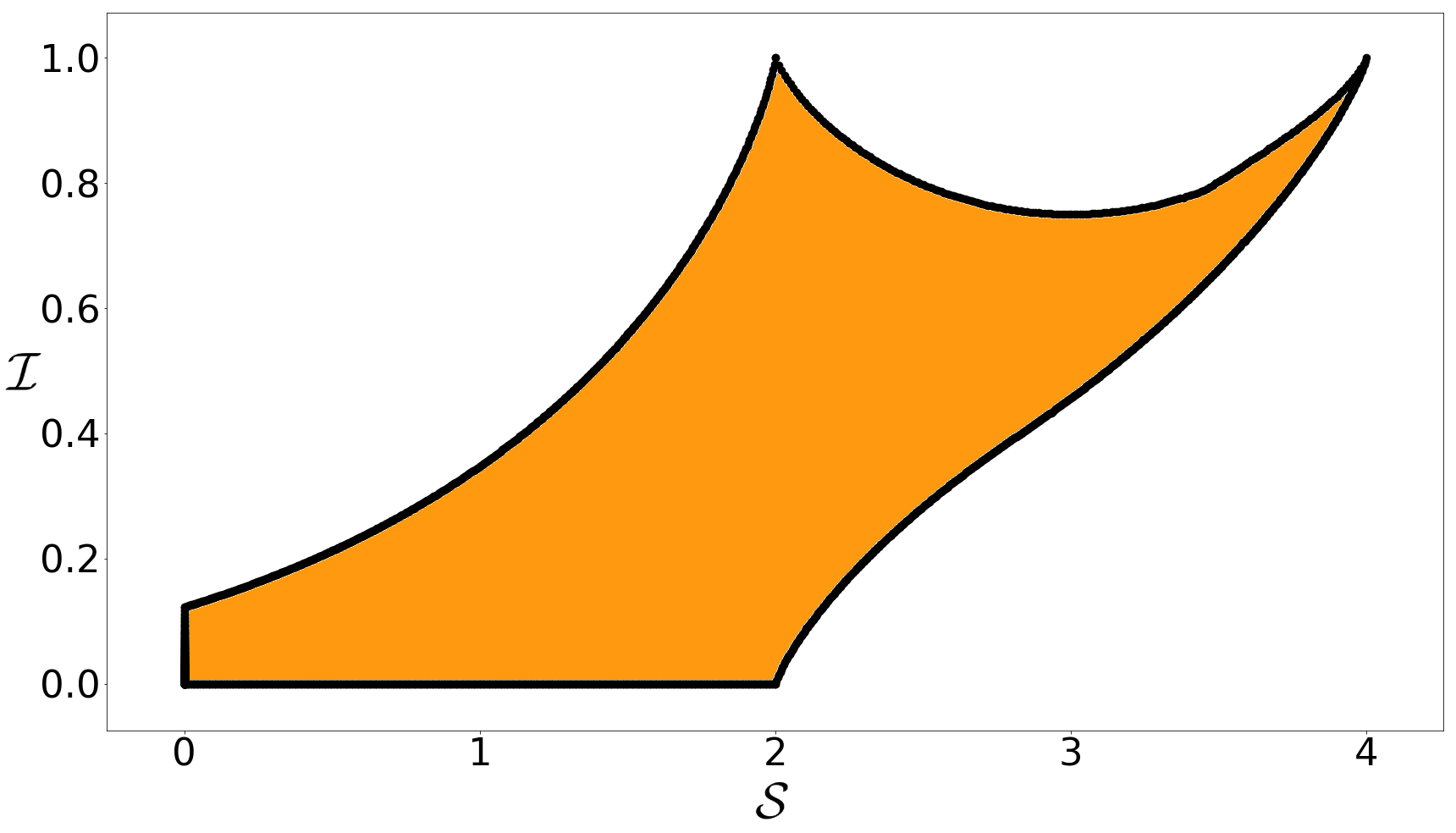}
    \caption{Boundaries of $\mathcal{NS}$ in the $\mathcal{S}-\mathcal{I}$ representation obtained by means of numerical optimization. The behaviors in $\mathcal{NS}$ fill the orange region on this representation.}
    \label{fig:bordesNS}
\end{figure}

Fig.~\ref{fig:bordesNS} shows the lower and upper bounds for $\mathcal{I}$ (that is, $I^{\mathcal{NS}}_{\mathit{max}}\left(\mathcal{S}\right)$ and $I^{\mathcal{NS}}_{\mathit{min}}\left(\mathcal{S}\right)$) obtained by means of numerical optimization, for $700$ values of $\mathcal{S}$ in $[0,4]$ for the maximums and $500$ values of $\mathcal{S}$ in $[2,4]$ for the minimums (we do not need to minimize $\mathcal{I}$ in the local region because it is straightforward to obtain analytically that $I^{\mathcal{\mathcal{L}}}_{\mathit{min}}\left(\mathcal{S}\right) \equiv 0$). 

The $\mathcal{S}-\mathcal{I}$ representation of $\mathcal{L}$, i.e.\ the $\mathcal{S}\in[0,2]$ region in Fig.~\ref{fig:bordesNS}, is pretty straightforward to analyze. For any fixed value of $\mathcal{S}\leq 2$, it is easy to find a behavior that factorizes as a product of its marginals: $p(ab|xy)=p(a|x)p(b|y)$ giving $\mathcal{I}=0$ and hence the lower bound for the local region. The upper bound for this region shows that  the maximum correlation that can be achieved is a monotonously increasing function of $\mathcal{S}$. 

The situation in the nonlocal region becomes more interesting: the upper bound for $\mathcal{I}$ turns immediately into a decreasing function once we leave the local set, showing that there is a compromise between this quantity and the magnitude of a Bell violation. As it is well-known, nonlocality does not 
necessarily mean stronger correlations. On the other hand, nonlocality cannot arise from product distributions and therefore the lower bound for $\mathcal{I}$ becomes a non-zero function of $\mathcal{S}$. This lower bound is a monotously increasing function of $\mathcal{S}$ over all the nonlocal region, and small values of $\mathcal{I}$ are less and less achievable as we move away from the corresponding CHSH facet of $\mathcal{L}$. A more detailed analysis of this curve leads to the most puzzling result of this work, but we will get into that later. In this respect, let us recall that no physical model was assumed when obtaining this plot, and the numerical optimization was done just imposing positivity, normalization and non-signaling constraints.

\subsection{$\mathcal{SYM}$ behaviors}\label{sec:sym}

Now we will see that, notably, the representation of $\mathcal{SYM}$ in the $\mathcal{S}-\mathcal{I}$ plane is exactly the same as that of $\mathcal{NS}$. This is a useful result that will simplify part of our analysis,  allowing us to obtain some analytic expressions for the boundaries of the non-signaling set in this representation. 

\begin{figure}
    \centering
    \includegraphics[width=1\columnwidth]{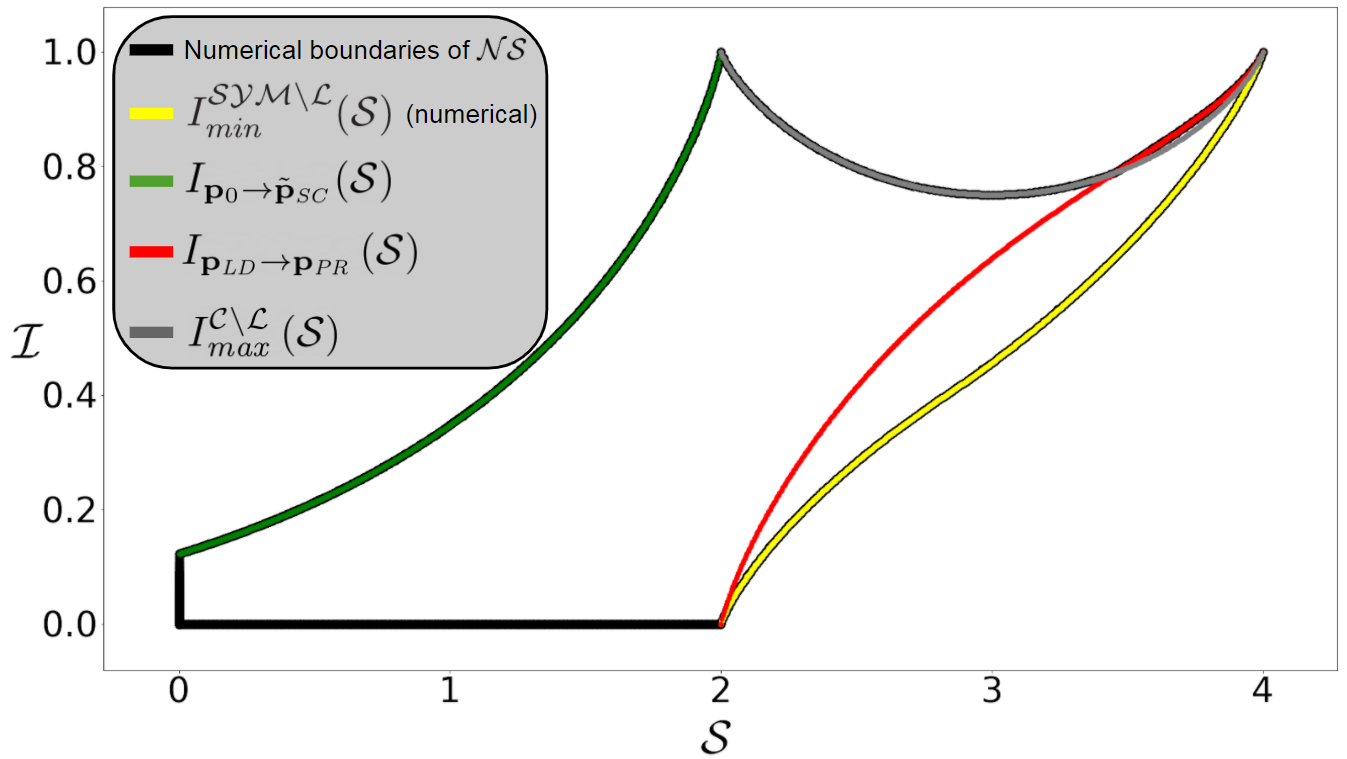}
    \caption{Boundaries of the $\mathcal{NS}$ set in the $\mathcal{S}-\mathcal{I}$ representation when restricting to behaviors in the $\mathcal{SYM}$ set. For comparison, in black we plot the points obtained by  numerical optimization  using the full set of $\mathcal{NS}$ behaviors.}
    \label{fig:bordesNSsimetrico}
\end{figure}

First, we will focus on the characterisation of $\mathcal L$. 
In this case, it is easy to check that a behavior maximizing $\mathcal{I}$ for $\mathcal{S}=0$ is given by  $\mathbf{p}_0 \in \mathcal{SYM}$ such that:
\begin{equation}
\vm{A_0}=\vm{B_0} = - \frac{1}{2} \; , \; \vm{A_1} = \vm{B_1} = \frac{1}{2} \, ,
\end{equation}
and all the other correlators equal to zero. Another extremal
point of this set, ${( \mathcal{S},\mathcal{I} ) = (2,1)}$, 
is reached by the following relabelling of the shared coin behavior:
\begin{equation}
\vm{A
_xB_y} = \begin{cases} -1 \; &\text{if} \; x=y \\
+1 \; &\text{otherwise}
\end{cases} \, ,
\end{equation}
and local mean values equal to zero. We will call this behavior $\tilde{\mathbf{p}}_{\mathit{SC}}\in\mathcal{SYM}$. In this way, the local upper bound for $\mathcal{I}$ 
can be obtained by evaluating all convex combinations between ${\bf p}_0$ and $\tilde{\mathbf{p}}_{\mathit{SC}}$. 
 Thus, the function:
\begin{equation}\label{eq:Lmaxanalitico}
I_{\mathbf{p}_0\rightarrow \tilde{\mathbf{p}}_\mathit{SC}} (\mathcal{S}) \equiv  \mathcal{I} \left( \frac{\mathcal{S}\tilde{\mathbf{p}}_{\mathit{SC}} + (2-\mathcal{S}) \mathbf{p}_0   }{2} \right) \, ,
\end{equation}
gives the local upper bound $I^{\mathcal{L}}_{\mathit{max}}\left(\mathcal{S}\right)$.  Fig. \ref{fig:bordesNSsimetrico} shows the agreement between this curve and the numerical results previously presented. 

For the nonlocal region, let us first note that when considering behaviors on the correlation space $\mathcal{C}$, mutual information takes a particularly simple form:
\begin{equation}\label{eq:info}
\mathbf{p} \in \mathcal{C} \Rightarrow \mathcal{I} [\mathbf{p}] = \sum_{xy} g\left( \vm{A_xB_y} \right) \, ,
\end{equation}
where we have defined $g:[-1,1] \rightarrow [0,\frac{1}{4}]$ as:
\begin{equation}
g(x) \equiv \frac{  1}{2}\left [1 + \frac{1+x}{4} \log \left( \frac{1+x}{4} \right) +\frac{1-x}{4} \log \left( \frac{1-x}{4} \right)\right],\nonumber
\end{equation}
which is an even function of its argument. 
It is clear that the suprema of $\mathcal{I}(\mathbf{p})$ with $\mathbf{p}\in\mathcal{C}$ is $1$ and is achieved if and only if $| \vm{A_xB_y}| = 1$ for all $x,y$; that is, for the shared coin behavior or the PR behavior, in agreement with the numerical optimization presented before. The condition 
$\left| \vm{A_xB_y} \right| = 1$ for all $x,y$ imposes also that $\mathcal{S}$ is either $2$ or $4$. Thus, crossing a CHSH facet comes at a price in terms of mutual information. As we will see later, this price is even bigger when we restrict ourselves to quantum behaviors.

In order to obtain the upper bound of $\mathcal{I}$ in the nonlocal part, one can consider behaviors
on $\mathcal{C}$.
Without loss of generality, we can stick to a particular labelling and set:
\begin{equation}\label{eq:Sp}
\mathcal{S}({\bf p}) = \vm{A_0B_0} + \vm{A_0B_1} + \vm{A_1B_0} - \vm{A_1B_1} \, .
\end{equation}
Then, we want to maximize  the $\mathcal{I}({\bf p})$ given in Eq.~\eqref{eq:info} subject to the linear constraint
$\mathcal{S}({\bf p}) =s$
for some $s\in[2,4]$. It is simple to show that the solution is attained when three of the correlators are
equal to $1$, and the remaining one is such that the constraint $\mathcal{S}({\bf p}) =s$ is satisfied. 
That is, the behavior with:
\begin{equation}
\vm{A_0B_0} = \vm{A_0B_1} = \vm{A_1B_0} = 1 \, , \, \vm{A_1B_1} = 3 - s \, , 
\label{eq:maxnolocalcorr}
\end{equation}
maximizes $\mathcal{I}$ over $\mathcal C$. Notice that for each $s \in [2,4]$, the behavior \eqref{eq:maxnolocalcorr} is not only an element of $\mathcal{C}$ but also of $\mathcal{SYM}$, and corresponds to a convex combination of the shared coin behavior and the PR behavior. Then, we have  that the curve:
\begin{equation}
I^{\mathcal{C}\setminus \mathcal{L}}_{\mathit{max}}\left(\mathcal{S}\right) = \mathcal{I} \left( \frac{(\mathcal{S}-2)\mathbf{p}_{\mathit{PR}} + (4-\mathcal{S}) \mathbf{p}_{\mathit{SC}}    }{2} \right), 
\label{eq:curvamaxnolocal}
\end{equation}
gives the maximum value of $\mathcal{I}$ of the behaviors in $\mathcal{C}$ space for $\mathcal{S}\in[2,4]$. The corresponding plot is also shown in Fig.~\ref{fig:bordesNSsimetrico}. This plot shows that there is a value of $\mathcal{S}$ from which the analytic curve \eqref{eq:curvamaxnolocal} starts deviating from the numerical upper bound. Fortunately, this ramification can also be obtained as the mutual information of a mixture of two simple behaviors. In fact, if we call $\mathbf{p}_{\mathit{LD}}$ the local deterministic behavior with $\vm{A_x} = \vm{B_y} = \vm{A_xB_y} = 1$ for all $x,y$, then the following function:
\begin{equation}
I_{\mathbf{p}_\mathit{LD} \rightarrow \mathbf{p}_\mathit{PR}}  \left(\mathcal{S}\right) = \mathcal{I} \left( \frac{(\mathcal{S}-2)\mathbf{p}_{\mathit{PR}} + (4-\mathcal{S}) \mathbf{p}_{\mathit{LD}}    }{2} \right), 
\label{eq:curvamaxnolocal2}
\end{equation}
 gives the upper bound for $\mathcal{I}$ in the nonlocal region after the intersection with \eqref{eq:curvamaxnolocal} (see in Fig.~\ref{fig:bordesNSsimetrico}). Notice that this is also the mutual information for a set of elements in $\mathcal{SYM}$. 

So far, we have derived the upper bound for $\mathcal{I}$ as a function of $\mathcal{S}$ for the non-signaling set. These bounds can be obtained by considering only symmetrical behaviors and  can be summed up as:
\begin{equation}\label{eq:NSmaxanalitico}
I^{\mathcal{NS}}_{\mathit{max}}\left(\mathcal{S}\right)\hspace{-0.3ex} = \hspace{-0.3ex}\begin{cases} I^{\mathcal{L}}_{\mathit{max}}\left(\mathcal{S}\right) &\text{if} \, \mathcal{S} \leq 2 \\
\max \left\{  I_{\mathbf{p}_\mathit{LD} \rightarrow \mathbf{p}_\mathit{PR}}  \left(\mathcal{S}\right) , I^{\mathcal{C}\setminus \mathcal{L}}_{\mathit{max}}\left(\mathcal{S}\right) \right\}  \hspace{-1.75ex} &\text{if} \, \mathcal{S} > 2
\end{cases}
\end{equation}

In order to show that all the boundaries of $\mathcal NS$, in the $\mathcal{I}-\mathcal{S}$ representation, can be obtained by considering only the symmetric part of this set, we have to consider also 
lower bounds. For the local part, this is straightforward because the segment $\mathcal{I}=0$ for $\mathcal{S} \in [0,2]$ is constructed from product behaviors, and this can be done within the symmetry assumption. The lower bound of the nonlocal part is non-trivial, and we do not have an analytic expression for the full set (as we will see later, we do have an analytic expression for a fraction of it). 
Thus, here we will just mention that the numerical minimization, when subjected to the additional constraint of symmetry, gives the same lower bound, $I_{\mathit{min}}^\mathcal{NS} \left( \mathcal{S} \right)$, as for the full set (see Fig.~\ref{fig:bordesNSsimetrico}). Finally, note that the fragment of $I_{\mathit{min}}^\mathcal{NS} \left( \mathcal{S} \right)$ lying on the post-quantum region (that is, for $\mathcal{S} \in [2\sqrt{2},4]$) can be also analytically described by:
\begin{equation}\label{eq:parteanaliticaminimos}
I_{\mathbf{p}_\mathit{Bell} \rightarrow \mathbf{p}_\mathit{PR}}  \left(\mathcal{S}\right) \equiv  \mathcal{I} \left( \frac{\left(\mathcal{S}-2\sqrt{2}\right)\mathbf{p}_{\mathit{PR}} + (4-\mathcal{S}) \mathbf{p}_{\mathit{Bell}}    }{4-2\sqrt{2}} \right) \, ,
\end{equation}
that is, by all the convex combinations between the Bell behavior and the PR behavior.

\subsection{The quantum set}\label{sec:quantum}

In order to characterize the quantum set we will consider the sets, $\tilde{\mathcal{Q}}$ and $\mathcal{Q}_1$, that contain the quantum set $\mathcal{Q}$.  The analytic expressions defining those two sets in this scenario,  can be used to find their boundaries in the $\mathcal{I}-\mathcal{S}$ representation.
First, note that constraints \eqref{eq:Qhatcond} cannot be satisfied by any behavior with $\mathcal{S} > 2\sqrt{2}$, so $\tilde{\mathcal{Q}}$ is contained in $\mathcal NS$. 
Given that $\mathcal{L}\subset \tilde{\mathcal{Q}}$, the boundaries in the $0\leq \mathcal{S}\leq 2$ region are those already obtained in the previous section. The last two observations imply that we only need to consider the boundaries of $\tilde{\mathcal{Q}}$ in the $\mathcal{S} \in [2,2\sqrt{2}]$ region. 
In addition, given that $\mathcal{Q}_1 \subset \tilde{\mathcal{Q}}$, this is also true for the set of behaviors belonging to the first level of the NPA hierarchy. Moreover, for $\mathcal{S} \in [2,2\sqrt{2}]$ we can check that all behaviors obtained through numerical minimization of $\mathcal I$, pass the test defining the first level of the NPA hierarchy. Thus, the lower bound for $\mathcal{Q}_1$ matches what we called $I_{\mathit{min}}^\mathcal{NS} \left( \mathcal{S} \right)$, if $2\leq \mathcal{S} \leq 2\sqrt{2}$. As $   \mathcal{Q}_1 \subset \tilde{\mathcal{Q}}\subset\mathcal{NS}$ this remains true for $ \tilde{\mathcal{Q}}$. 

\begin{figure}
    \centering
    \includegraphics[width=1\columnwidth]{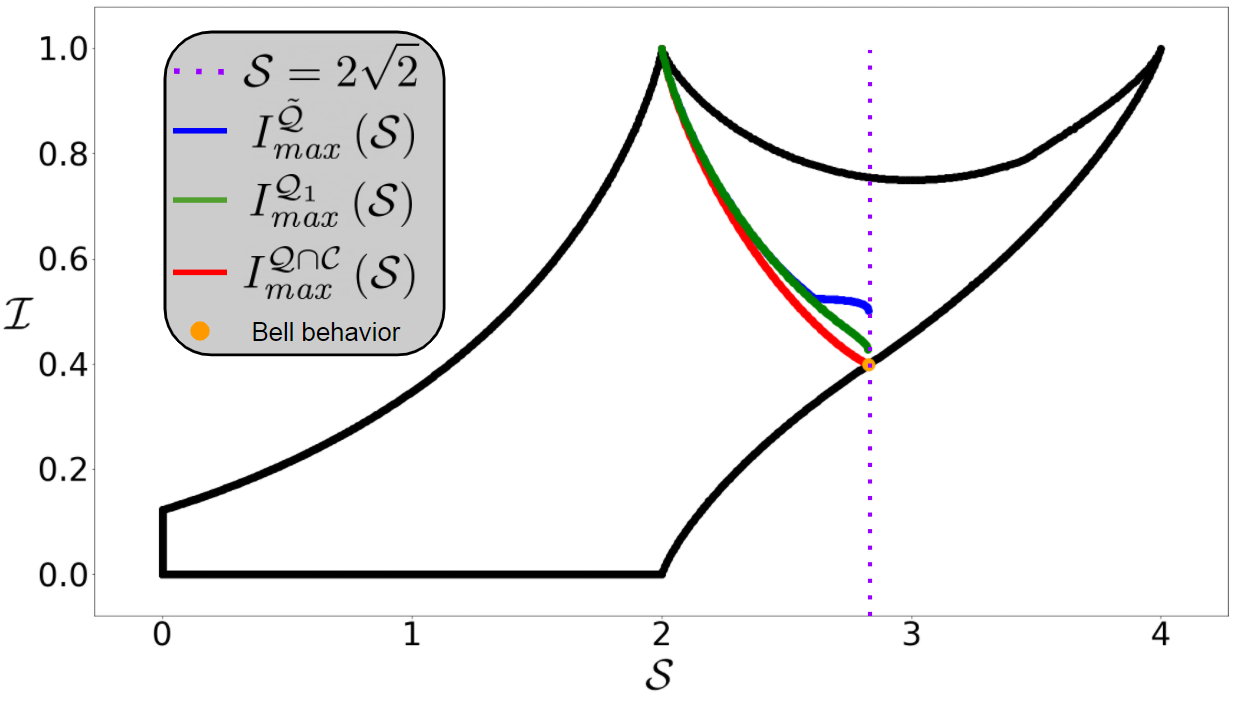}
    \caption{Boundaries for the $\mathcal NS$ set and the quantum sets. Upper bounds of $\mathcal I$: $I^{\tilde{\mathcal{Q}}}_{\mathit{max}}\left(\mathcal{S}\right)$ in blue; $I^{\mathcal{Q}_1}_{\mathit{max}}\left(\mathcal{S}\right)$ in green; and $I^{\mathcal{Q}\cap \mathcal{C}}_{\mathit{max}}\left(\mathcal{S}\right) $ in red. The lower bound for these sets is the same as the one for $\mathcal NS$, the orange point represents the Bell behavior.}
    \label{fig:bordesQ}
\end{figure}

As we mentioned before, for behaviors in $\mathcal{C}$ the set of relations \eqref{eq:Qhatcond} are not only necessary but also sufficient conditions to be in $\mathcal{Q}$. Thus, we will show that
it is  possible to find the upper bound analytically for behaviors in $\mathcal{C}$. We want to maximize
$\mathcal{I} (\mathbf{p})$ in Eq.~\eqref{eq:info} but, in this case, subjected to the set of non-linear inequalities Eqs.~\eqref{eq:Qhatcond}. As before, we set $\mathcal{S}({\bf p})=s$ (Eq.~\eqref{eq:Sp}) for $s\in [2,2\sqrt{2}]$. Given that the maximum value of the CHSH is reached by that labelling, then the constraints in Eq.~\eqref{eq:Qhatcond} reduce to just one inequality: $\arcsin  \vm{A_0B_0}  + \arcsin \vm{A_0B_1} 
+  \arcsin  \vm{A_1B_0}  - \arcsin \vm{A_1B_1}  \leq \pi $. Therefore, the upper bound for $\mathcal{I}$ over the set $\mathcal{Q} \cap \mathcal{C}$ is given by:
\begin{equation}
I^{\mathcal{Q}\cap \mathcal{C}}_{\mathit{max}}\left(\mathcal{S}\right) = 3g\left( w(\mathcal{S}) \right) + g\left(3 w(\mathcal{S}) - \mathcal{S} \right), 
\label{eq:curvamaxQC}
\end{equation}
where
\begin{equation}
 w (\mathcal{S}) = \sqrt{2} \cos \left[  \frac{\pi}{6} + \frac{1}{3} \arctan\left( \frac{\mathcal{S}}{\sqrt{8-\mathcal{S}^2}} \right) \right].
\end{equation}
Fig.~\ref{fig:bordesQ} shows the boundaries obtained by means of numerical optimization for $\mathcal{Q}_1$ and $\tilde{\mathcal{Q}}$. Note that the strict inclusion $\mathcal{Q}_1 \subset \tilde{\mathcal{Q}}$ is preserved in this representation and the same holds for $\mathcal{Q} \subset \mathcal{Q}_1$.
This is so, given that the Bell behavior is the only quantum behavior with $\mathcal{S}=2\sqrt{2}$ and, thus, the lower and upper bound of this set have to intersect at this value of $\mathcal{S}$, which is not true for $\mathcal{Q}_1$. The figure also shows the analytic curve \eqref{eq:curvamaxQC} for the upper bound of $\mathcal{Q}\cap\mathcal{C}$. It can be seen that, as expected, this curve intersects the lower bound exactly at the Bell behavior. All these boundaries show that the relation between mutual information and the magnitude of a Bell violation is indeed stronger for quantum behaviors than for arbitrary non-signaling behaviors.

\begin{figure}
    \centering
    \includegraphics[width=1\columnwidth]{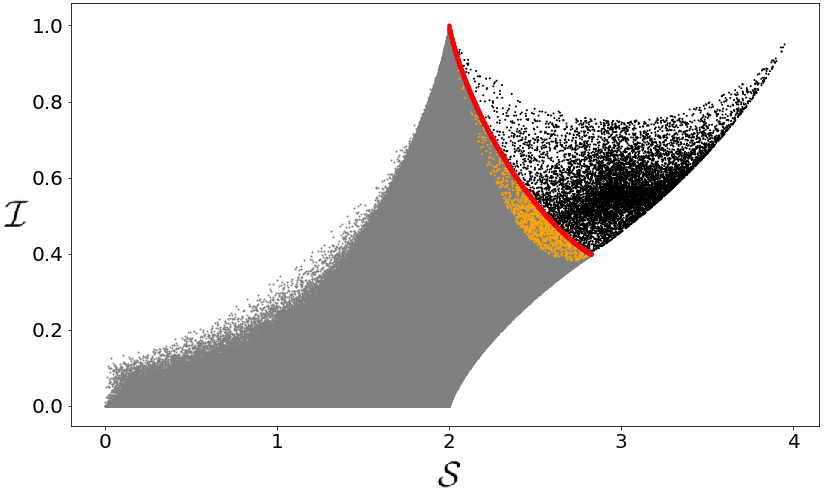}
    \caption{Random sample of behaviors. 
    Gray points are $5\times10^6$ quantum behaviors sampled for Hilbert spaces of 
    dimension two, it can be seen that all of them lie below the red line, representing the upper bound for
    the set  $\mathcal{Q}\cap\mathcal{C}$. The other points are random convex combinations of $\mathbf{p}_{\mathit{SC}}$, $\mathbf{p}_{\mathit{Bell}}$ and $\mathbf{p}_{\mathit{PR}}$: behaviors not passing the  test for $\tilde{\mathcal{Q}}$ are shown in black, while for the rest are in orange. }
    \label{fig:bordeQc}
\end{figure}

In order to perform another test of these bounds, in Fig.~\ref{fig:bordeQc} we show the sampling of $5 \times 10^6$ random quantum behaviors (gray points) of the form \eqref{eq:quantbehpurif} with Hilbert spaces of dimension $2$. There it is also shown the analytic curve \eqref{eq:curvamaxQC} upper bounding the set $\mathcal{Q}\cap \mathcal{C}$. Additionally, it is shown the sampling of $10^4$ random points in $\mathcal{C}$ obtained as a mixture bewteen the shared coin,  Bell  and  PR behaviors:
the orange (black) points do (do not) pass the test of Eq.~\eqref{eq:Qhatcond}.
There we can see more explicitly how \eqref{eq:curvamaxQC} is the proper bound for the quantum behaviors in the correlation space. 

\subsection{The Tsirelson bound}\label{sec:tsirelson}

Perhaps the most puzzling result of this work is the natural appearance of the Tsirelson bound in this representation.
If we look carefully to the lower bound of the mutual information in the nonlocal region, $I^{\mathcal{NS}}_{\mathit{min}}\left(\mathcal{S}\right)$,  we can identify, in the neighborhood of the Bell behavior, a change in the concavity of the curve. Lets try to find out exactly where it happens. To avoid working with second derivatives, which can carry a lot of numerical noise, we can take advantage of the following trick: given an ordered set of three points in a two-dimensional plane, $\vec{A} = (x_A,y_A)$, $\vec{B} = (x_B,y_B)$ and $\vec{C} = (x_C,y_C)$; the orientation of the path $\vec{A}\rightarrow\vec{B}\rightarrow\vec{C}$ is related to the sign of the determinant of the following matrix \cite{wikideterminante}:
\begin{equation}
\mathbf{O} = \begin{bmatrix}
1 & x_A & y_A \\
1 & x_B & y_B \\
1 & x_C & y_C \end{bmatrix}. 
\label{eq:determinante}
\end{equation}
The points are orientated clockwise if $\det(\mathbf{O}) < 0$ and counterclockwise if $\det(\mathbf{O}) > 0$. Therefore, we can take the numerical values of $\mathcal{I}^{\mathcal{NS}}_{\mathit{min}}\left(\mathcal{S}\right)$ to evaluate the concavity of the curve at any value of $\mathcal S$ using the determinant of the matrix in Eq.~\eqref{eq:determinante}.

%

In order to get a detailed inspection of the region where the concavity change occurs, we obtain again (through numerical optimization) $I^{\mathcal{NS}}_{\text{\textit{mín}}}\left(\mathcal{S}\right)$ for $5000$ equispaced values of $\mathcal{S}$ in the interval $[ 2.5 , 3.1 ]$ (note that the Tsirelson bound lies in this interval, given that $2\sqrt{2} \simeq 2.83$). For what comes next, it will be useful to give a name to the horizontal separation between the points in this curve, $\delta \mathcal{S} \equiv \frac{3.1 - 2.5}{5000} $.
 
\begin{figure}[H]
    \centering
    \includegraphics[width=1\columnwidth]{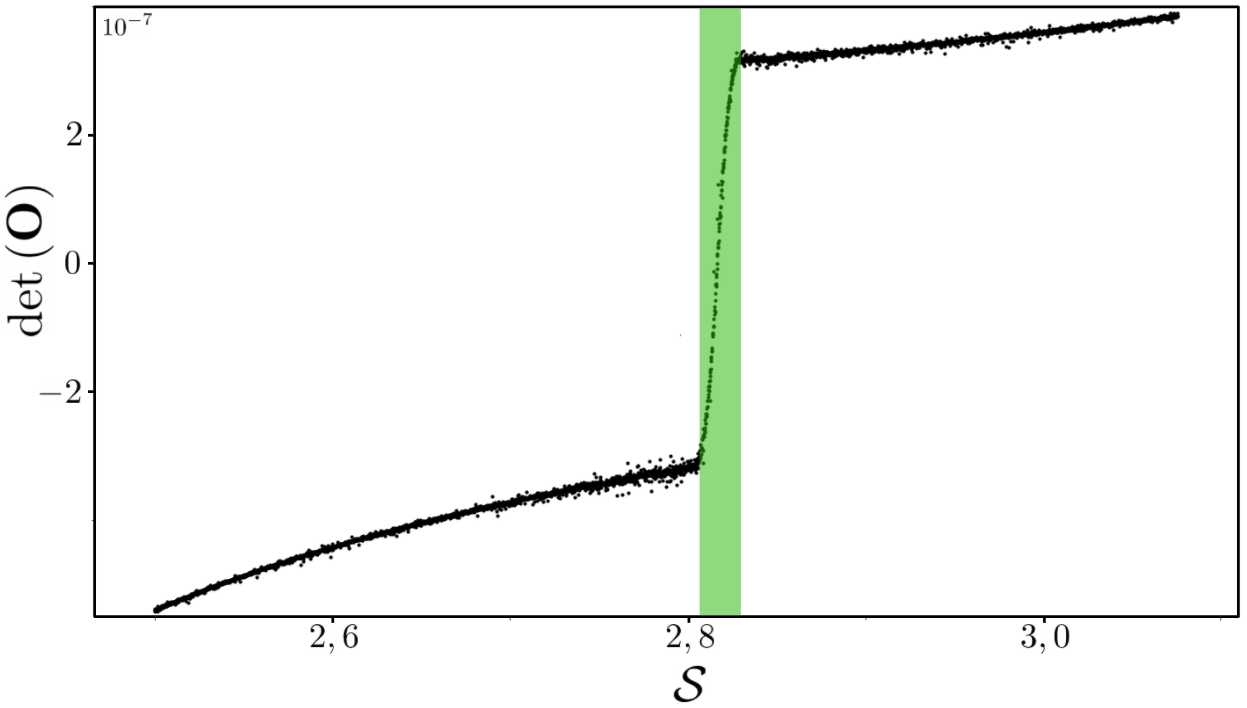}
    \caption{Determinant of $\mathbf{O}$ for sets of three points in the curve $I^{\mathcal{NS}}_{\text{\textit{mín}}}\left(\mathcal{S}\right)$. The horizontal axis gives the value of $\mathcal{S}$ for the first point of each set. The green highlighted area shows those points where $\det \left( \mathbf{O} \right)$ does not allow us to decide, numerically, whether the curve is concave or convex.}
    \label{fig:concavidad1}
\end{figure} 

Fig. \ref{fig:concavidad1} shows, for each value of $\mathcal{S}$, the determinant of $\mathbf{O}$ for the corresponding point and two other points to the right of it. To avoid excessive numerical noise, it is convenient not to take three consecutive points, because in that case $\mathbf{O}$ is almost singular and computing its determinant becomes tricky from a numerical point of view. For that reason, the determinant computed for each value of $\mathcal{S}$ corresponds to the point with that value of $\mathcal{S}$, the point with $\mathcal{S} + 100 \delta \mathcal{S}$ and the point with $\mathcal{S} + 200 \delta \mathcal{S}$. Therefore, in this plot, we should expect to find a region of width $200 \delta \mathcal{S}$ where the sign of $\det \left( \mathbf{O} \right)$ does not have a valid interpretation in terms of concavity, because this region (highlighted in green) corresponds to values of $\mathcal{S}$ where the curve is still concave but in the matrix $\mathbf{O}$ we are taking into account points where the curve is already convex. From the last observation, it follows the way in which we should interpret where the concavity change happens, and it is in the value of $\mathcal{S}$ from which the transition region ends and the plot of $\det \left( \mathbf{O} \right) $ as a function of $\mathcal{S}$ stabilizes to the new curve of positive values.

\begin{figure}
    \centering
    \includegraphics[width=1\columnwidth]{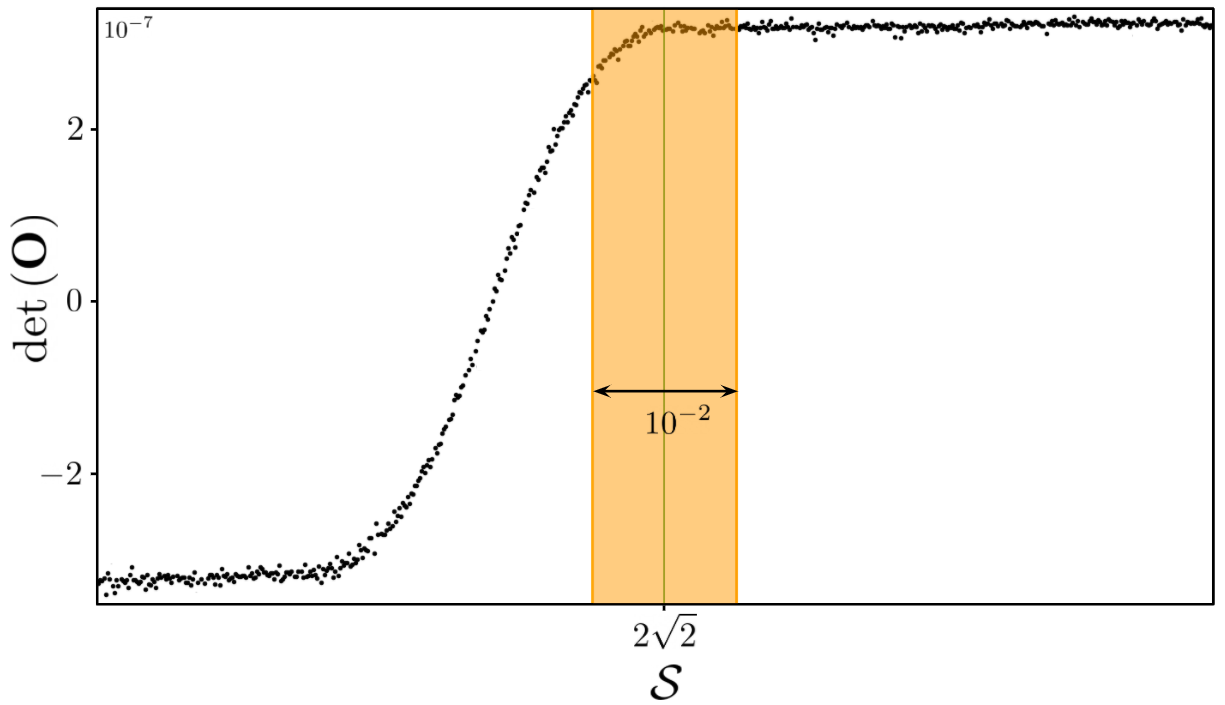}
    \caption{Determinant of $\mathbf{O}$ as a function of $\mathcal{S}$ in the neighborhood of the concavity flip. The vertical green line gives the value of the Tsirelson bound. The change on the concavity of $I^{\mathcal{NS}}_{\text{\textit{mín}}}\left(\mathcal{S}\right)$ occurs inside of the orange region.}
    \label{fig:concavidad2}
\end{figure}

Fig. \ref{fig:concavidad2} shows the detail of the mentioned transition region. The vertical green line indicated the value of the Tsierlson bound and the orange region is a symmetrical neighborhood of size $10^{-2}$ around it. We see that the end of the transition region (and, hence, the value of $\mathcal{S}$ where the concavity flips) is located in the range $\mathcal{S} = 2\sqrt{2} \pm 0.005$. The error in the last expression can be made smaller by taking an increasing amount of points when obtaining the lower bound for the mutual information around the Bell behavior, so this result gives robust numerical evidence that the inflection point is, indeed, located at the Tsirelson bound.

\begin{figure}
    \centering
    \includegraphics[width=1\columnwidth]{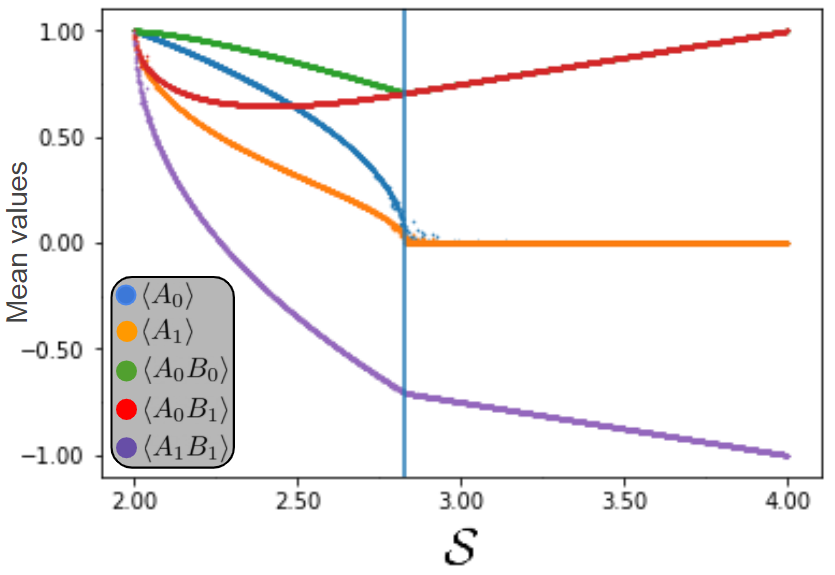}
    \caption{Values of the five relevant mean values for symmetrical behaviors along the curve
     $I^{\mathcal{NS}}_{\mathit{max}}\left(\mathcal{S}\right)$ in the $\mathcal{S}\in[2,4]$ region. 
    The vertical line is located at $\mathcal S=2\sqrt 2$. }
    \label{fig:valoresmedioscurvamin}
\end{figure}

At the Tsirelson bound we can also find a singularity when looking at the behaviors that give place to the curve $I^{\mathcal{NS}}_{\text{\textit{mín}}}\left(\mathcal{S}\right)$. As we have shown, there is no loss of generality in describing this curve by taking into account only symmetrical behaviors, so five parameters are enough to describe any of its points. Fig.~\ref{fig:valoresmedioscurvamin} shows the values of the five correlators that correspond to the
behaviors along $I^{\mathcal{NS}}_{\mathit{max}}\left(\mathcal{S}\right)$ when going from $\mathcal{S}=2$ to $\mathcal{S}=4$. Once the Bell behavior is reached, both local mean values stay fixed to zero and the correlators vary lineally from the values corresponding to the Bell behavior to those corresponding to the PR behavior, in agreement with \eqref{eq:parteanaliticaminimos}. As we can see, the parametrizations of the mean values in terms $\mathcal{S}$ show discontinuous changes in their first derivatives when going through the Bell behavior.


In this section we have shown robust numerical evidence that the curve $I^{\mathcal{NS}}_{\text{\textit{mín}}}\left(\mathcal{S}\right)$ changes its concavity  at $\mathcal{S} = 2\sqrt{2}$ and, furthermore, the associated curve of behaviors is not smooth, because it exhibits singularities in its first derivatives at the Bell behavior. Those two facts are rather intriguing, given that the curve $I^{\mathcal{NS}}_{\mathit{min}}\left(\mathcal{S}\right)$ is found without making use of quantum mechanics at all. Therefore, this result might suggest that the Tsirelson bound can be obtained in a device-independent manner. 

\section{Summary}\label{sec:summary}

In this work we presented an alternative approach to describe the set of non-signaling correlations that is based on a two dimensional representation of the non-signaling polytope, combining a geometrical quantity
(the maximum value of a Bell functional) with an informational one (mutual information). 
Bell functionals provide us with a kind of measure of the nonlocal content of each behaviour, while
the mutual information is a measure of the total correlations between the parties. 
We showed that there exists a trade-off between mutual information and the magnitude of the maximum Bell violation once the nonlocal region is reached, and this trade-off is more restrictive for quantum behaviors. 
Our analysis also suggests that the Tsirelson bound can be obtained from a device-independent argument, as we showed that in this scenario it appears as an inflection point in the curve for the lower bound of the mutual information. An interesting question that we leave for future analysis is  
whether this approach of finding quantum bounds from the study of informational related quantities 
can provide further insights in more general Bell scenarios.

\section*{Acknowledgement}

This work was supported by ANPCyT (grants PICT 2014-3711, 2015-2293, 2016-2697 and 2018-04250), CONICET-PIP 11220130100329CO and UBACyT.

\bibliography{Biblio}

\end{document}